\documentclass[conference]{IEEEtran}
\IEEEoverridecommandlockouts
\usepackage{cite}
\usepackage{amsmath,amssymb,amsfonts}
\usepackage{algorithmic}
\usepackage{graphicx}
\usepackage{textcomp}
\usepackage{xcolor}
\def\BibTeX{{\rm B\kern-.05em{\sc i\kern-.025em b}\kern-.08em
    T\kern-.1667em\lower.7ex\hbox{E}\kern-.125emX}}
\begin{document}

\title{Geo-Spatial Data Visualization and Critical Metrics Predictions for Canadian Elections\\
%%{\footnotesize \textsuperscript{*}Note: Sub-titles are not captured in Xplore and should not be used}
%% \thanks{Identify applicable funding agency here. If none, delete this.}
}

\author{\IEEEauthorblockN{Mohammad Abdul Hadi}
\IEEEauthorblockA{\textit{Department of Computer Science} \\
\textit{University of British Columbia}\\
British Columbia, Canada \\
hadi@alumni.ubc.ca}
\and
\IEEEauthorblockN{Fatemeh Hendijani Fard}
\IEEEauthorblockA{\textit{Department of Computer Science} \\
\textit{University of British Columbia}\\
British Columbia, Canada \\
fatemeh.fard@ubc.ca}
\and
\IEEEauthorblockN{Irene Vrbik}
\IEEEauthorblockA{\textit{Department of Computer Science} \\
\textit{University of British Columbia}\\
British Columbia, Canada \\
irene.vrbik@ubc.ca}
}

\maketitle

\begin{abstract}
Open data published by various organizations is intended to make the data available to the public. All over the world, numerous organizations maintain a considerable number of open databases containing a lot of facts and numbers. However, most of them do not offer a concise and insightful data interpretation or visualization tool, which can help users to process all of the information in a consistently comparable way. Canadian Federal and Provincial Elections is an example of these databases. This information exists in numerous websites, as separate tables so that the user needs to traverse through a tree structure of scattered information on the site, and the user is left with the comparison, without providing proper tools, data-interpretation or visualizations. 

In this paper, we provide technical details of addressing this problem, by using the Canadian Elections data (since 1867) as a specific case study as it has numerous technical challenges. We hope that the methodology used here can help in developing similar tools to achieve some of the goals of publicly available datasets. The developed tool contains data visualization, trend analysis, and prediction components. The visualization enables the users to interact with the data through various techniques, including Geospatial visualization. To reproduce the results, we have open-sourced the tool. 
\end{abstract}


\section{Introduction}
Open data is a movement to make data accessible for public use, intending to allow people to manipulate the data (e.g., using software tools) for, among others, linking datasets, mapping, and visualizations \cite{Gurstein}. 
Canada is one of the leading countries toward this effort, with Open Parliament as an example of making data available for public and data enriched citizen engagement in policy \cite{parliament}.
The effective use of open data requires many considerations, including technical challenges, interpreting the data, and availability of visualizations \cite{Gurstein}.
Although a massive amount of data is available and is being maintained by the government as well as independent organizations in the form of web-based databases, concise and insightful data-interpretation using adequate visualization techniques is found to be lacking \cite{Slocum}, \cite{openData}. 
There are multiple efforts towards making publicly available data usable, such as DBPedia, to extract structured information from Wikipedia \cite{Auer}, and narrative visualization of Swiss open data \cite{openData}. 

Visualizing the data can be helpful in the interpretation of the results\cite{openData}, which can be powered by data analytics and prediction models. However, there are rare works that integrate the visualizations with more advanced techniques such as trend analysis and prediction components in one place. Moreover, when the Geo-spatial data is not included in the metadata of the tables, plotting the data points on maps is a challenge, as extracting and demonstrating the correct information is difficult. Therefore, in this paper, we provide details of the tool we developed to collect, visualize, and analyze open data. We made the tool available as open-source\footnote{https://github.com/Mohammad-Abdul-Hadi/scraper-for-canadianelectionsdatabase.ca}.

We use the Canadian Federal and Provincial Elections data as a case study, as it contains various smaller tables and numerous branches that a user needs to traverse. The dataset includes the Election results since 1867 and is challenging to interpret as separate tables. Therefore, it can be a valuable case of publicly open data that cannot be used efficiently.  

\textit{Motivation of the case study.}
The 2019 Canadian Federal Election took place on October 21st, 2019, for electing members of the House of Commons to the 43rd Canadian Parliament. As a part of the election campaign, participating parties distributed a large number of pamphlets and leaflets among all the voters. But for making an informed decision, the voters must know the outcomes and details of the past elections. Voters use different websites to gather data about number of candidates, party popularity over time, or the geographic trends of party support for any given election and further analyze the data for trends. To best of our knowledge, no data-interpretation or visualization tool has been developed so that voters can gather all this information effortlessly.

\textit{Objective.} To fulfill our objective, we intend to design an interactive platform that would provide a Geospatial visualization of Canadian provinces and color-code them according to the party-wise election outcomes in any chosen election year. The Canadian map representing election data would also be accompanied by some powerful graphs to improve users' comprehension about the past elections and help them see underlying trends for the forthcoming elections. Users would be provided options from all the available federal and provincial elections over the past 152 years. 

We also intend to integrate an auxiliary component in our tool to predict some important metrics for future election by using the available data from the past elections. Moreover, the technical details and the approach that we provide in this paper can help to develop similar tools and overcome some challenges in interpreting the open data for non-expert users. 

\textit{Contribution.} We deliver a visualization tool where we have used the Choropleth map to show the outcome of elections on specific election year with colored geographical regions according to the associated winner party in that region. Users are also able to explore the massive amount of election data in the most efficient way where all correlated factors or metrics like “Seats won by Political Parties,” “Votes shared by Political Parties,” “Seats shared by Political Parties” for any given election would be categorically and graphically represented. 

For the election data, we have relied on a web-embedded database created by The School of Public Policy of University of Calgary where we can get all the relevant data for all the elections since 1867. 

The rest of this paper is organized as follows. In section II, the methodology and the architecture of the tool are explained, followed by Results of User Experiments and Discussions and Conclusions in Sections III and IV.

\section{Methodology and Tool Architecture}
The architecture of the tool consists of three main components. Each component requires different technologies to accomplish a given job. The functions of these components are explained below:

\textit{A. Python Scraper:}    We designed a python scraper that to efficiently scrape through the web-embedded database and store all the information in distinct comma-separated values. These files are the input of the other parts of the tool.

\textit{B. Geo-spatial Map Visualization:}    An R script that integrates various packages and libraries to enable the visualization of the Geospatial map of Canada. It assures that The maps represent proper election data that was gathered by the scraper.

\textit{C. Trend Analysis Component:} Tableau is used to generate different graphs for the user to comprehend the information better. This component also provides forecasting and trend analysis for certain metrics, i.e., the number of candidates, number of seats won by a certain party. Different options are provided for the users so that each user has a choice to see various representations of the same information and visualize different graph-types (e.g., horizontal/vertical bar-chart, pie/donut-chart, and heat-maps).

In the following, we describe the details of these three components.

\subsection{\textbf{Python Scraper}}

\begin{figure}[t]
    \centering
    \includegraphics[width=\linewidth]{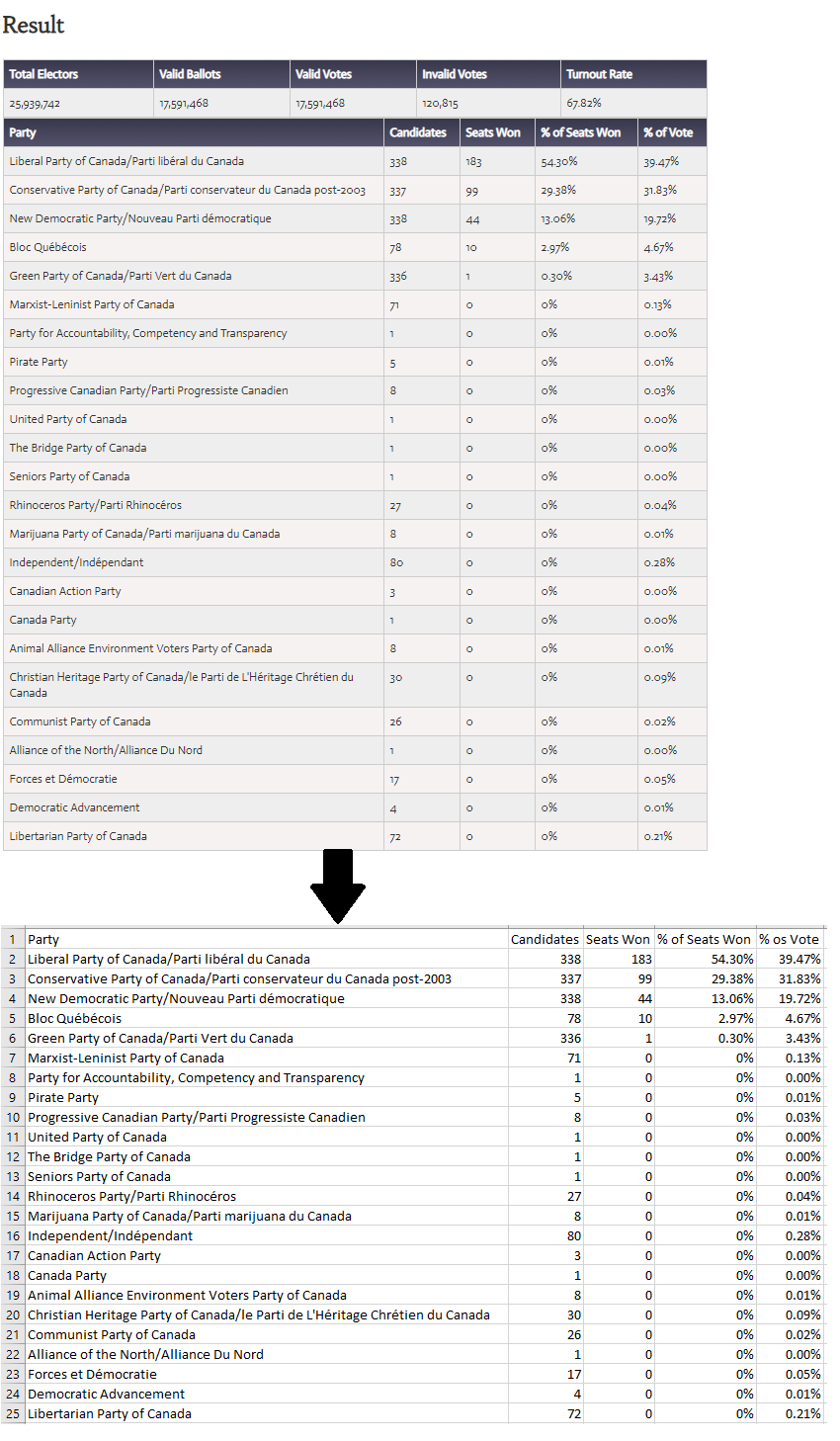}
    \caption{View of web-embedded Database to \textit{.csv} file extraction (static pages)}
    \label{fig:viewDB}
\end{figure}

For this case study, we require to extract validated dataset of Canadian elections from different resources. The database that is developed by Dr. Anthony Sayers at the Department of Political Science at the University of Calgary is one of the most reliable, consistent, and accessible databases for Canadian elections \cite{Sayers}. The details of the database are given here:
http://canadianelectionsdatabase.ca/
This database contains information on federal, provincial, and territorial elections since 1867 and is arranged by-election, party, candidate, and district. The database allows users to explore the data in numerous ways. 
We have chosen this database as it contains around four million data points \cite{Sayers}, and represents the difficulties of interpreting open data for users. 

\textit{A challenge} of working with this data is the lack of having a \textit{download option} to retrieve and use this information (e.g., import to analysis tools). Therefore, we developed a scraper to scrape, extract necessary data, and store them in an appropriate format, to reduce further data cleaning and polishing, whereas storing data as it is found in the database may lead to exhaustive data-formatting for the re-usability purpose. The extracted data is intended to be fed to the other two components.
An example of one of the tables on this data (from the database website) is shown in the top part of the Fig. \ref{fig:viewDB}. After the whole process of scraping, we are going to acquire all the data stored in the mentioned web-embedded tables and save them as \textit{.csv} files (shown in the bottom part of Fig. \ref{fig:viewDB}) so that the data can be easily accessed later for the use of data interpretation or visualization tool.

\begin{figure}[t]
    \centering
    \includegraphics[width=\linewidth]{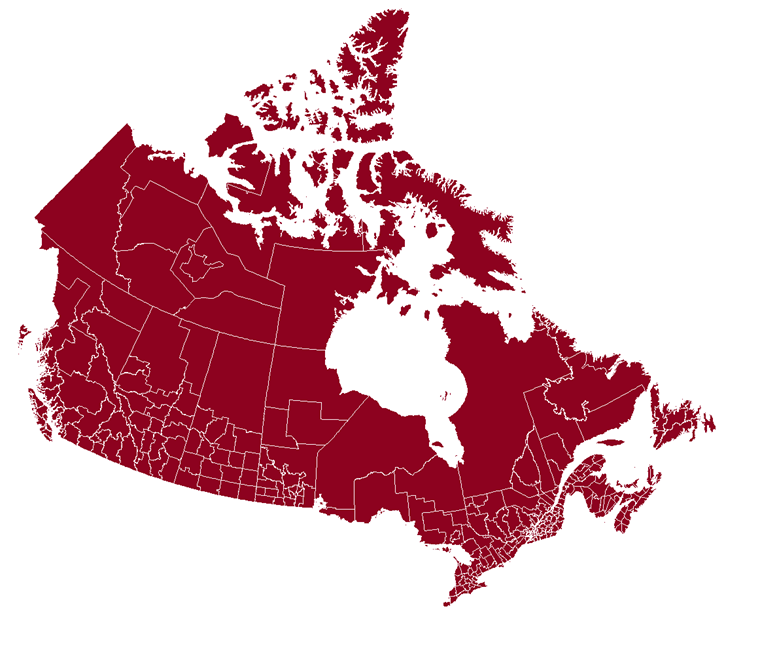}
    \caption{Basic Map of Canada (without color-coding)}
    \label{fig:Fig4}
\end{figure}

We have chosen Python to develop the scraper component as it provides a powerful, robust package for web-crawling and data scraping, namely "Beautiful Soup." Beautiful Soup is a Python package for parsing HTML and XML documents (including having malformed markup, i.e., non-closed tags, so named after tag soup). It creates a parse tree for parsed pages that are used to extract data from HTML web pages \cite{Richie}.
\textit{Beautiful Soup} package provides methods to return/download the HTML page upon sending a request to the server with the corresponding URL. From the downloaded HTML page, we can search, identify, and retrieve required components (document segments) of the page such as "table" using their designated class or id. Once we locate the required element, the data is parsed into a \textit{list} to be stored later in a \textit{.csv} file. The intermediate \textit{list} helps to modify huge chunks of data as per the requirement of other components in the project (for convenience and reusability). The scraped election data is stored as census-district-level data and province-territory-level-data, shown in the top part of Fig. \ref{fig:viewDB}. We have named different tables storing different election information in the following format "$\langle election-type \rangle$ \_ $\langle year\rangle$" where \textit{election-type} refers to either Federal or Provincial election and \textit{year} refers to the year when the election took place.  

\begin{figure}[t]
    \centering
    \includegraphics[width=\linewidth]{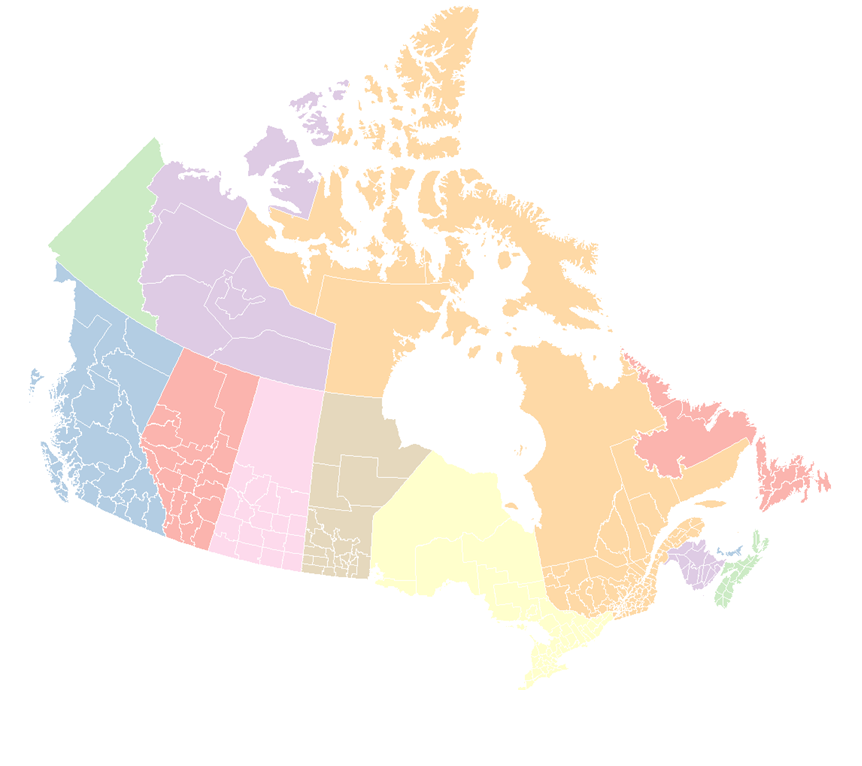}
    \caption{Basic Map of Canada (with color-coded provinces)}
    \label{fig:Fig5}
\end{figure}

\textit{A particular challenge} was that some pages were communicating with the database using dynamic AJAX and JS requests. Beautiful Soup does not provide support for such scenarios. These pages were inspected separately for the specific requests, which would, in turn, return the data that we require. 
To retrieve the data that is fetched by AJAX or JS request, our code listens on the port that communicates with the database and downloads the database response. To accomplish this task, we utilized another powerful, robust package \textit{Selenium} that enabled us to capture all the AJAX and JS responses.
Lastly, these methods are highly dependent on the "connection-time." If the response is not received within a specified period, the method terminates listening to the port. 
A try-catch block captures the method and converts the connection-time to be an incrementing one-hot-encoding variable to address this issue. 
As a result, for a specified connection-time, if the method can not capture a request, it would try again with an increased connection-time and returns to base value after each loop.

The static and dynamic pages are handled differently, as explained above. Each program extracts and stores the retrieved data separately.

\subsection{\textbf{Geo-spatial Map Visualization}}

For this part of the project, we have developed a Choropleth map (i.e., a map in which we shaded the areas in proportion to a statistical variable) of Canada and integrate it with the election information. The gist of the process is grabbing a \textit{.shapefile} (geospatial vector data format for geographic information system software) and converting it to simple-features objects to be used in R.

The \textit{tidyverse} is used as the core package to convert map data into a plot. Converting map data into a format (that R packages can use) requires a lot of different technical steps as they cannot be used directly with the provided methods in \textit{tidyverse} libraries. Other packages: \textit{sf, rgdal, geojsonio, spdplyr, rmapshaper} are also used, which provide functionalities for conversion and mapping process.
As part of this process, we have built a separate function, \textit{theme\_map()}, as a ggplot theme that turns off insignificant pieces of the plot (so that it looks neat).

In the next step, the actual map data (\textit{.shapefile}) of different regions is extracted for developing a Canadian map. We collected the data from the Canadian Central Statistics Agency: http://www12.statcan.gc.ca/census-recensement/2011/geo/bound-limit/bound-limit-2011-eng.cfm. 
This Shape File format \textit{(.shp)} is the most popular and widely-used standard format for map data.
In our tool, we selected \textit{ArcGIS .shp} files as they contain the category Census divisions and cartographic boundaries, and it is convenient for the integration of the election data. 
These shapefiles are imported in our \textit{R} script as an object using the \textit{readOGR} function of \textit{rgdal} package and then are converted into GeoJSON format to simplify the polygons. This GeoJSON file becomes the building block for further components. After that, we read the GeoJSON file back as an \textit{sf (simple features)} object. 

These steps of data importing, converting, and thinning take a long time to execute, the resulting data are saved for further use and are made available for the use of other researchers. the \textit{.geojson} the file is also included in the repository, so for the test purposes, one can load the data and start working form this point.

\begin{figure}[t]
    \centering
    \includegraphics[width=\linewidth]{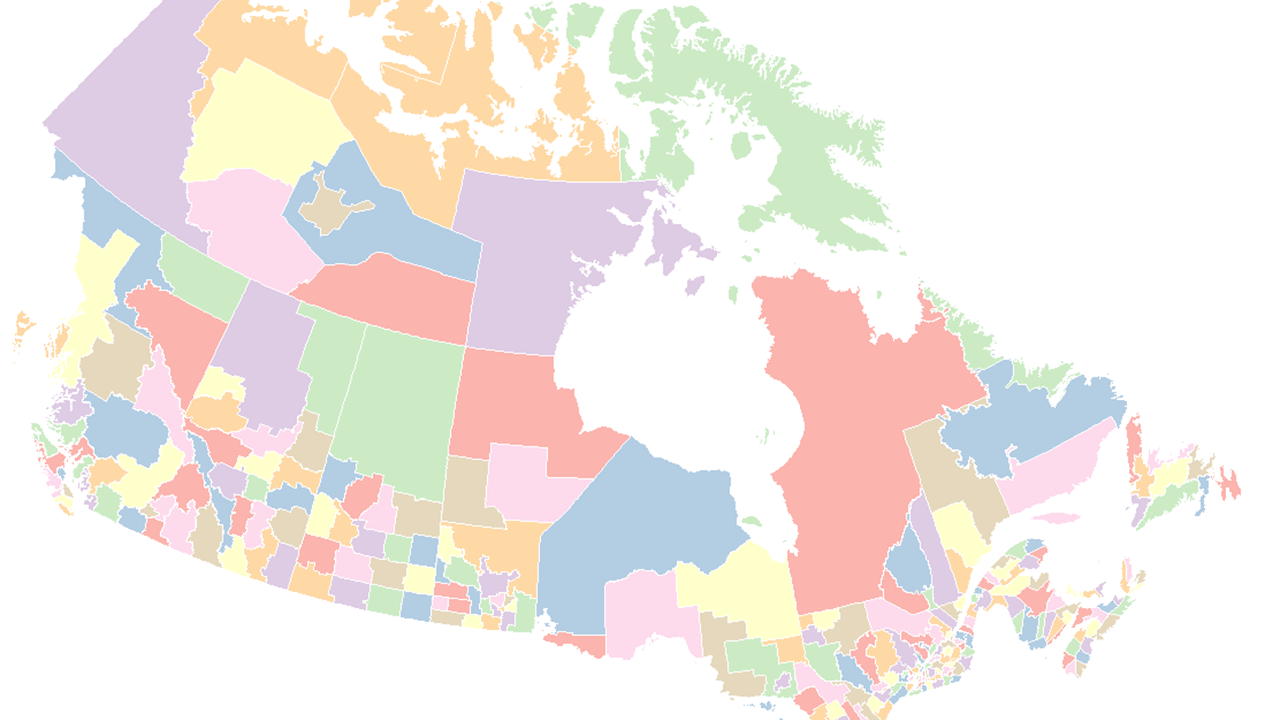}
    \caption{Basic Map of Canada (with color-coded census-district)}
    \label{fig:Fig6}
\end{figure}

\begin{figure}[b!]
    \centering
    \includegraphics[width=\linewidth]{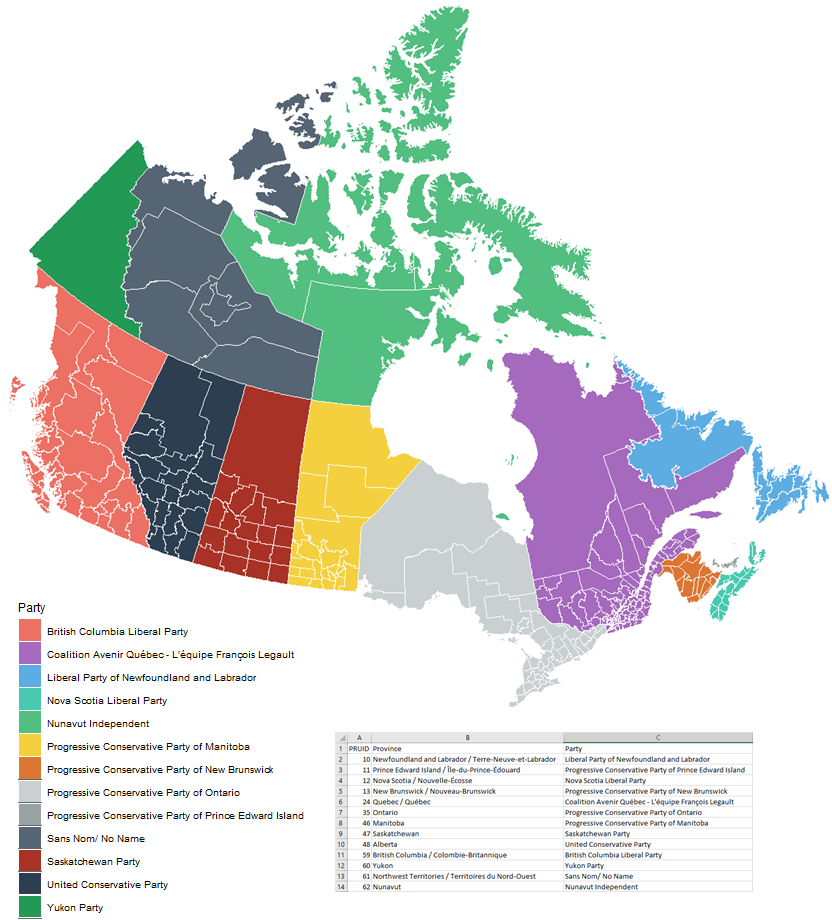}
    \caption{View of web-embedded Database (static pages)}
    \label{fig:map}
\end{figure}

Fig. \ref{fig:Fig4}, \ref{fig:Fig5}, \ref{fig:Fig6} represent the generated maps without the merging of the election data. The first part, Fig. \ref{fig:Fig4} shows the map with our putting any color code throughout the region; the second part, Fig. \ref{fig:Fig5} shows the map with color-coded provinces and the third part, Fig. \ref{fig:Fig6} shows a color-coded district throughout the Canadian region.

\begin{table}[t!]
\begin{center}
\caption{Corresponding Election Data extracted from Web embedded Database for Map generated in Figure \ref{fig:map}}
\label{table:electionData}
\begin{tabular}{|p{.83cm}|p{3cm}|p{3cm}|} 
\hline
PRUID & Province & Party\\
\hline
48 & Alberta (Apr 16, '19) & United Conservative Party\\
59 & British Columbia (May 9, '17) & British Columbia Liberal Party\\
46 & Manitoba (Apr 19, '16) & Progressive Conservative Party of Manitoba\\
13 & New Brunswick (Sep 24, '18) & Progressive Conservative Party of New Brunswick\\
10 & Newfoundland and Labrador (May 16, '19) & Liberal Party of Newfoundland and Labrador\\
12 & Nova Scotia (May 30, '17) & Nova Scotia Liberal Party\\
62 & Nunavut (Oct 30, '17) & Nunavut Independent\\
61 & Northwest Territories (Oct 3, '11) & Sans Nom/ No Name\\
35 & Ontario (Jun 7, '18) & Progressive Conservative Party of Ontario\\
11 & Prince Edward Island (Apr 23, '19) & Progressive Conservative Party of Prince Edward Island\\
24 & Quebec (Oct 1, '18) & Coalition Avenir Québec - L'équipe François Legault\\
47 & Saskatchewan (Apr 4, '16) & Saskatchewan Party\\
60 & Yukon (Oct 1, '11) & Yukon Party\\
\hline
\end{tabular}
\end{center}
\end{table}

Finally, the scraped election data stored as census-district-level data and province-territory-level-data are converted into data frames and merged separately with the map-data produced as a \textit{sf} (simple- feature) object, namely \textit{canada-cd} (this is essentially a big data frame specifying a large number of lines that need to be drawn on the plot).  

This merging step required careful attention while matching the key variables to avoid introducing missing values; otherwise, the lines on the map would not have smoothly joined. If missing values are introduced, it would have resulted in a shredded map as R tries automatically to fill the missing portion of the polygons \cite{Healy}.

That is why it is worth mentioning that for plotting data on maps, joining two datasets on the character attributed column should be avoided \cite{Healy}. An example is provided to illustrate the issue. If we had taken \textit{PRNAME} (province name) column to merge the data, it would introduce \textit{null} values in the resulting- data frame as rows corresponding to the province "British Columbia" is referred as "British Columbia (BC)/Colombie britannique" in the election data frame but "British Columbia (BC)" in the corresponding map data frame. Broken Maps are usually caused by these kind of merge errors. Another example can be, one of the province names could contain a leading or trailing space as a result of data- extraction limitations (like "Alberta" and "Alberta" which would cause the join to fail). 
Therefore, another attribute is used to merge the data (in our case province id- PRID) as it has a numeric value.
Other issues that we have considered for plotting maps is converting the numeric columns back to factors and using specific data type (discrete or continuous) to avoid receiving errors.

\begin{figure}[t]
    \centering
    \includegraphics[width=\linewidth]{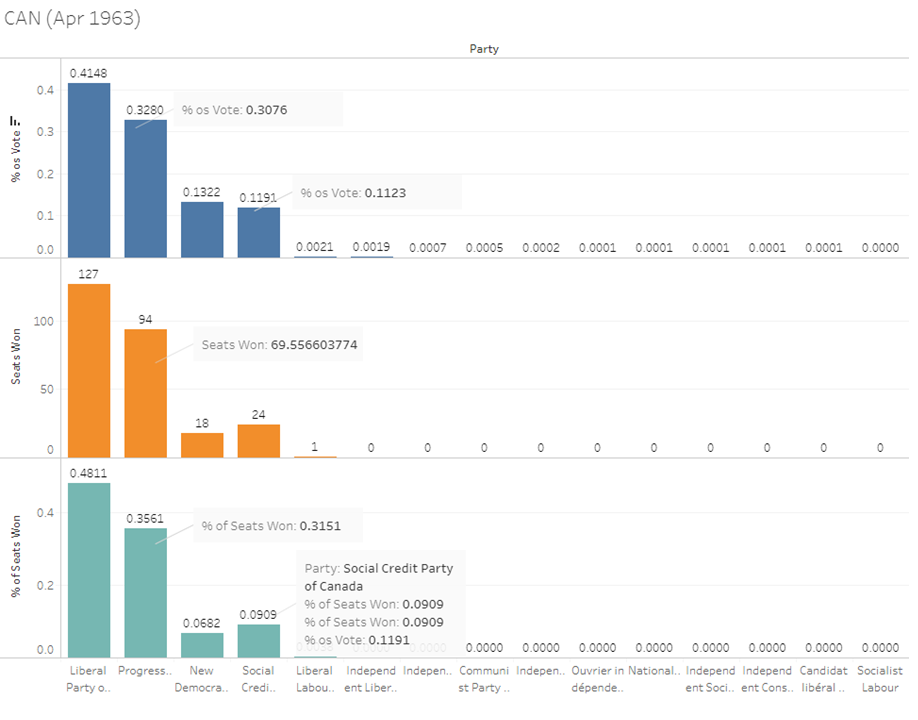}
    \caption{Information about Canadian Federal Election in 1963 is represented in Vertical Bar-chart}
    \label{fig:illustration1}
\end{figure}

Once the joining is complete, our R script generates the map with conventional \textit{ggplot} by filling the PRID (Province ID) for representing federal elections or the CDUID (Census-District ID) for provincial elections. Our interactive platform lets the user choose any election (federal or provincial) and the year. We just select a specific table storing the specified information, merge the table with map-data, and generate the Choropleth map for visualization.

For demonstration purposes, we have randomly selected one map (for the sake of brevity) that colors the regions based on the winner parties in the latest Province and Territorial election and shown the representation in Fig. \ref{fig:map}. This figure is produced from data shown in Table \ref{table:electionData}. At the bottom of this figure, all the party names are assigned different colors and then different regions of the Geospacial map are painted with the same color of the associated party won in that particular region. From the bottom right corner of the figure, we have attached the associated table-data that was retrieved from the web-embedded database.

Please note that this map is selected randomly from over 400 maps generated from the scraped data (31 elections for each of the 13 provinces). 

\subsection{\textbf{Trend Analysis Graphs}}

\begin{figure}[t]
    \centering
    \includegraphics[width=\linewidth]{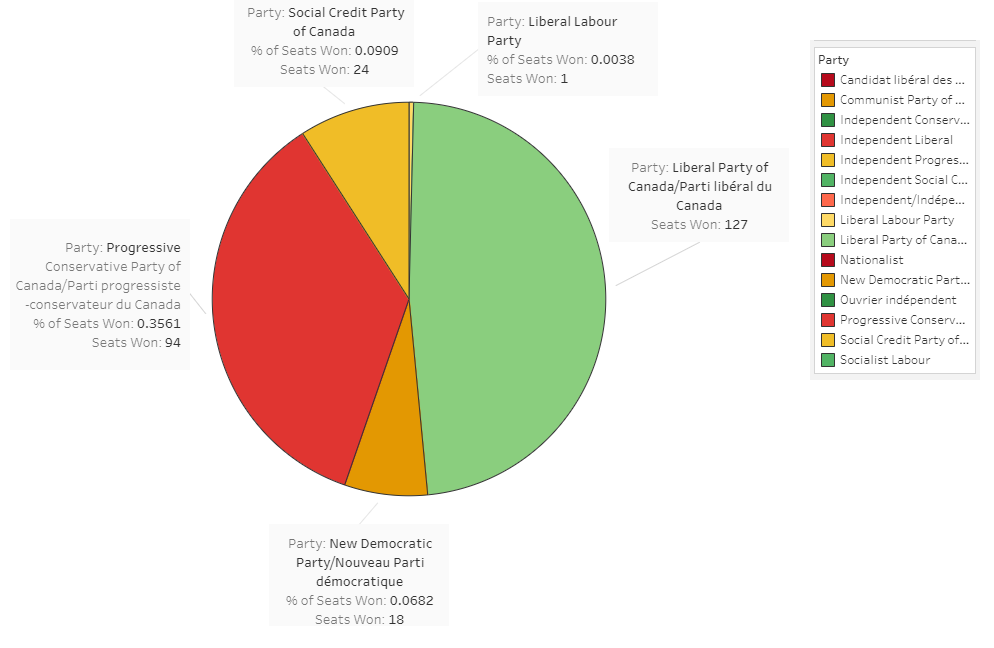}
    \caption{Information about Canadian Federal Election in 1963 is represented in Pie-chart}
    \label{fig:illustration2}
\end{figure}

\begin{figure}[b!]
    \centering
    \includegraphics[width=\linewidth]{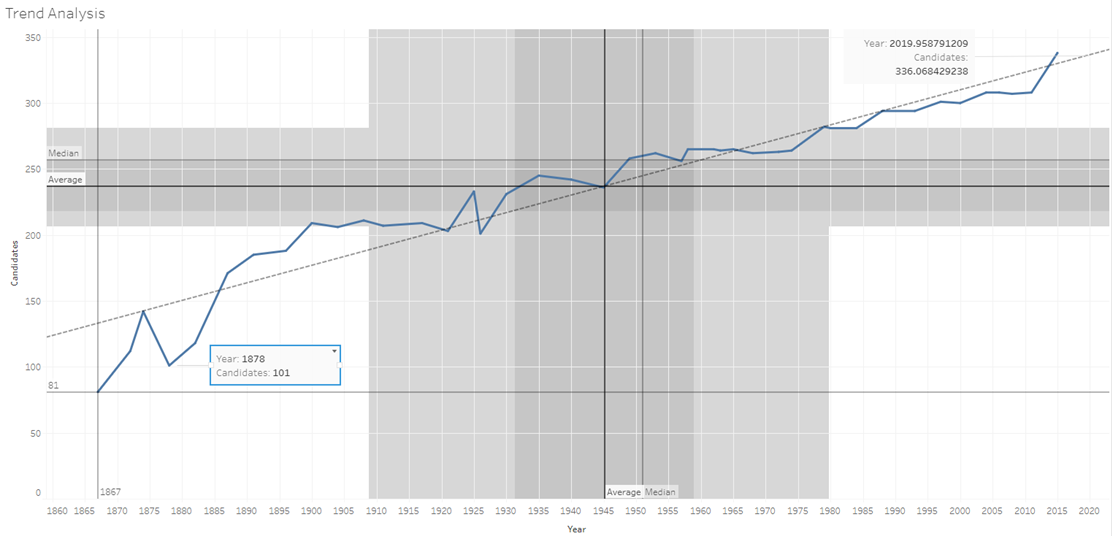}
    \caption{Total candidate numbers in the federal elections since 1867}
    \label{fig:canNum}
\end{figure}

Visualizing election-data on maps is helpful, but it cannot provide a lot of the required information and analysis. We developed the last component of our tool architecture to mitigate the map limitations and provide precise predictions on particular metrics for both federal and provincial elections \cite{keller}. We have used Tableau to generate different graphs of users' choice.

A demonstration on how user can choose from different data-interpretation or plotting techniques is provided in Fig. \ref{fig:illustration1} and \ref{fig:illustration2}. The user can choose any of these illustrations from the platform. The mentioned figures represent the Canadian Election data from 1963.  In Fig. \ref{fig:illustration1}, we have used a vertical bar chart to interpret 3 important metrics; "percentage of seats won," "the number of votes won," "percentage of seats won." The same metrics are also represented by the pie chart in Fig. \ref{fig:illustration2}. Using these representations, any user can quickly get the gist of the whole election and can easily identify the most and the least prominent players in the election. We do acknowledge the fact that every user would not be comfortable using the same representation. Keeping that in mind, in addition to these two representations, we have provided users lots of other conventional plotting techniques such as horizontal bar chart, and donut chart, scatter plot, and line plot.
From our iterative user study, we understood that providing different options for the users to explore the data is essential, as the data has different aspects to it and also ranges quite a lot.

\begin{figure}[t!]
    \centering
    \includegraphics[width=\linewidth]{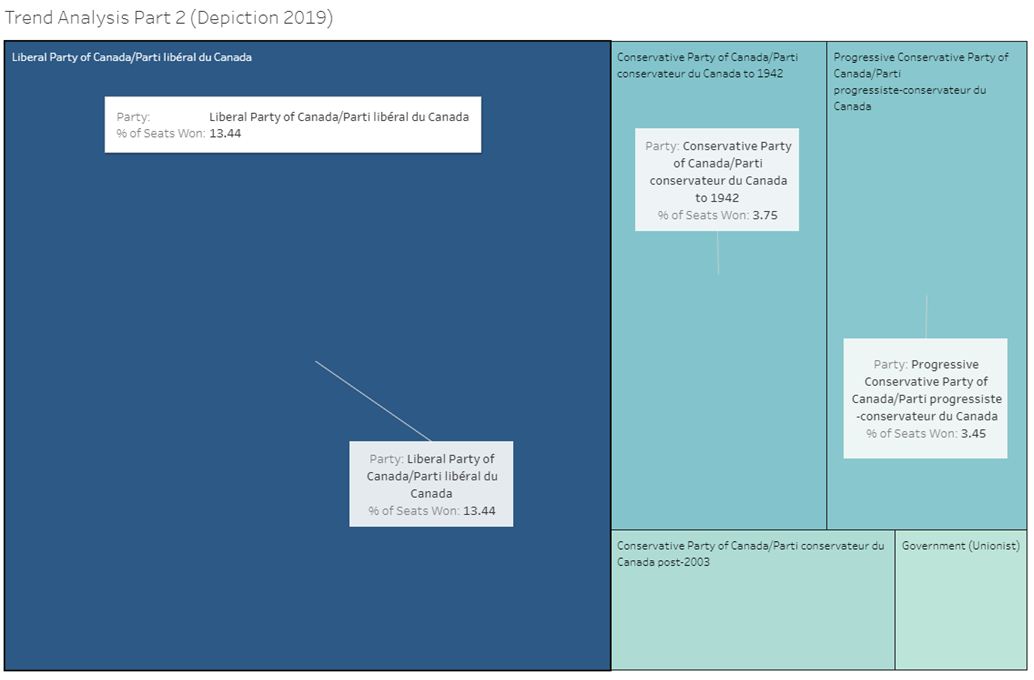}
    \caption{Federal parties’ history of winning in the past elections}
    \label{fig:heat1}
\end{figure}

The next illustration in Fig. \ref{fig:canNum} shows the total candidate numbers in the federal elections since 1867. We can see that the number of candidates is ever increasing. We have implemented a simple \textit{Linear Regression} algorithm to predict the number of candidates in the 2019 election \cite{Smola}. Prediction of the illustration (Candidate Number in 2019 election: ~336) is close to the original result. Here, we have also added information about the median and average of the number of candidates. If anyone wants details of an election since 1867, the user can hover the mouse over the specific point on the shown line, and the text-box pops up with all the interesting information about that election. For future predictions, the user can hover over any point on the best-fit line.

The illustration in Fig. \ref{fig:heat1} shows the federal parties' history of winning in the past elections. We have chosen heat-map as it has previously been used to compare the volumetric difference using color intensities. Therefore, a user can easily grasp the information like the following: "In the history of Canada, a specific party has won the most election" This is a review we got from our user during the testing phase of the tool. \cite{bias}. Users can also pick that the second and third most popular parties in Canada always went neck-to-neck when it comes to winning federal elections, although the user may need to possess certain knowledge about the party-names and which parties are still in play. But users can gain this knowledge from other graphs that are discussed in the project.

\begin{figure}[t!]
    \centering
    \includegraphics[width=\linewidth]{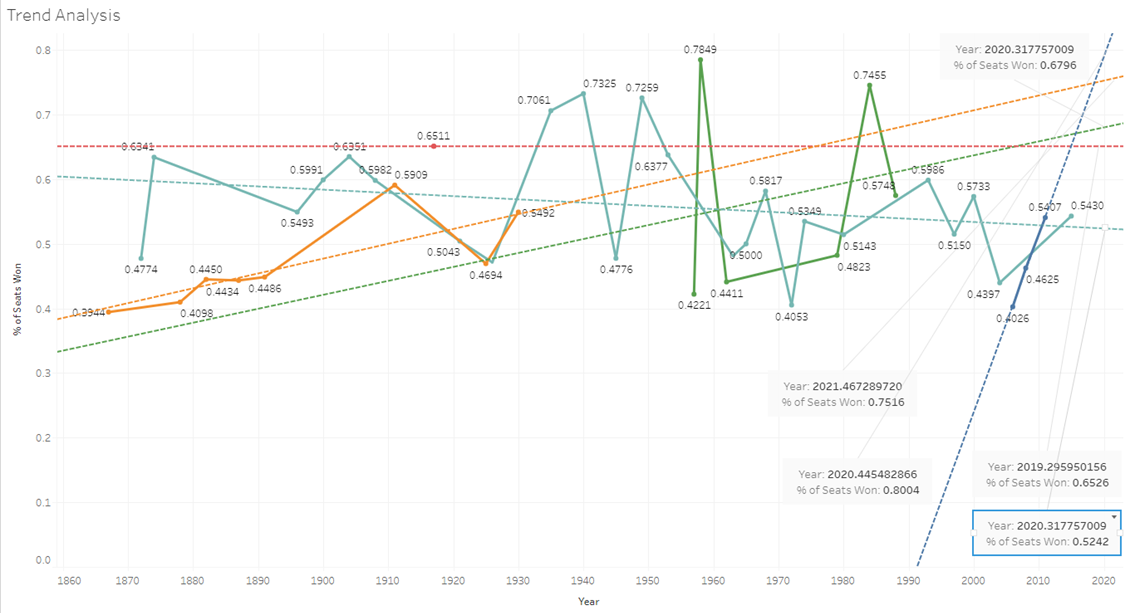}
    \caption{Merged representation for 3 metrics (number of seats won, percentage of seats won, percentage of votes won)}
    \label{fig:all}
\end{figure}

From our initial study, users revealed that having only year-wise or election-wise data representation is not helpful enough. Our users have requested a progression map where all the practical data should be present and accessible by the users at their convenience. In Fig. \ref{fig:all}, a representation is shown where the information of all of the parties is put together (e.g. "number of seats won," "percentage of seats won," "percentage of votes won") along with the regression lines generated for each party. This graph can identify which party has been consistent over the year, which party has been trending over the past couple of elections, which parties' popularity (a feature that is depicted by other attributes like percentage of seats and votes won) \cite{eppler}. For example, from this illustration, we can see that although the Liberal Party has won most of the elections throughout history, its' popularity has been slowly decreasing over the long run. In contrast, the Conservative Party of Canada is gaining tremendous support in recent decades, and Unionists have been holding their position steadily throughout the history \cite{taylor}.

\begin{figure*}[t]
    \centering
    \includegraphics[width=\linewidth]{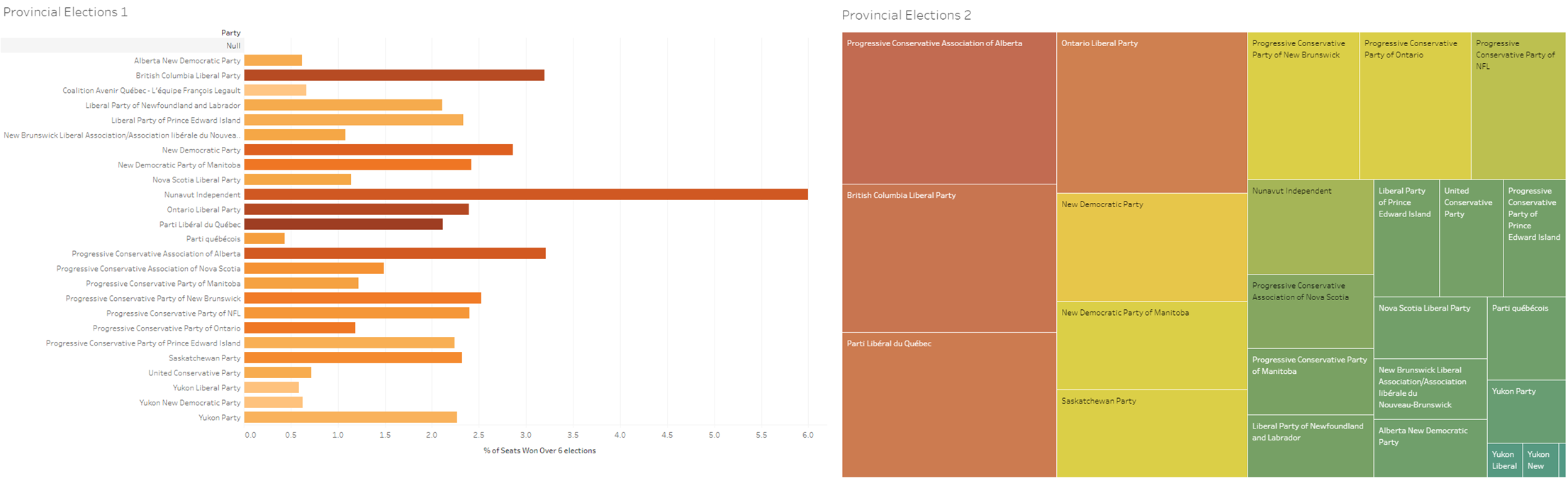}
    \caption{Provincial parties' gained most of the votes throughout the history}
    \label{fig:mf3}
\end{figure*}

The first part of Fig. \ref{fig:mf3} deals with provincial elections, unlike the other graphs. Here all the provincial parties participated to demonstrate which ones gained most of the votes throughout the history. The same information has been interpreted using the heat-map in the second part of \ref{fig:mf3}, where the user has the independence to choose from any representation.

\section{User Experiments' Results}

Although we are not reporting directly on the usability results of the study, we have tested the tool with ten users. The goal was to introduce the user to a data visualization tool that would interpret the Canadian election data in a sensible, comprehensible, reproducible way using different graphs and visualization methods. We have received positive feedback from the ten users who used our tool for gathering information about the Canadian election. Moreover, we recruited a small sample population (14 people) from the grad and undergrad students at random as users (there were 6 Canadian and 8 international/immigrant students). All of them were paid for giving us feedback upon the use of the tool. Six of these international students had almost no knowledge about Canadian Election. They were asked to use the tool for 10 minutes and to surf around the database for another 10 minutes. All of them reported that the visualization tool was far more effective for gathering knowledge, comprehension, and comparison. All of the 14 users preferred the visualization tool compared to scraping the raw data provided in the web-embedded database.

\textbf{Source Codes and Graphs.} To encourage replicability, we uploaded all scripts, codes, and graphs to the following link and provides the annotated dataset upon request \footnote{https://gitlab.com/Mashuk/ieee-ccece-geo-spatial-data-visualization-and-critical-metrics-predictions-for-canadian-elections}.

\section{%Discussions and 
Conclusion}
Open data is meant to be used by the public and provide data-driven decisions. However, visualization tools are required to make the data interpretable. One of the other challenges to provide the analysis and visualization tools for open data is the different formats of the data that are published by various parties in separate databases. 
In this paper, we provided architecture and the technical details of an open-source tool that we developed for collecting data, and visualizing and analyzing information. 
Although the tool is developed explicitly for Canadian Election data, the technical details and the approach can be used by researchers from various fields and developers to address the issue of open data (i.e., having separate databases with no interpretation tool). 

As a continuation of the work, we will run an empirical study to acquire user feedback for usability studies. Furthermore, we acknowledge that many other machine learning approaches can be adopted to predict numerous other metrics of the Canadian Election. Separately we have been developing and implementing different algorithms, but the benchmark result is yet to be determined.

\section{Acknowledgment}

This work is supported by NSERC Grant 05175.

\end{document}